\documentclass[numberedappendix]{emulateapj}
\newcommand{\mode}{emulateapj}
\usepackage{ifthen}
\newcommand{\mr}[1]{\mathrm{#1}}

\newcommand{\m}{$^{-1}$}

\newcommand{\kms}{\mr{km}\,\mr{s}^{-1}}
\newcommand{\Mpc}{\mr{Mpc}}

\ifthenelse{\isundefined{\nonaas}}{}{

\bibpunct{(}{)}{;}{a}{}{,}

  \renewenvironment{thebibliography}[1]{%
    \begin{oldthebibliography}{#1}%
      \setlength{\parskip}{0ex}%
      \parindent 0ex%
      \setlength{\itemsep}{0ex}%
  }%
  {%
    \end{oldthebibliography}%
  }

}

\usepackage{ifpdf}

\usepackage{graphicx}

\newcommand{\usemyrefs}{1}

\shortauthors{Mahdavi et al.}
\shorttitle{Line-of-Sight Extent of Clusters}
\ifthenelse{\equal{\mode}{aastex}}
{

}{
\journalinfo{}
\submitted{Submitted March 25, 2011; accepted May 11, 2011 for publication in ApJL}

}

\newcommand{\fone}{
\begin{figure}
\resizebox{3.5in}{!}{\includegraphics{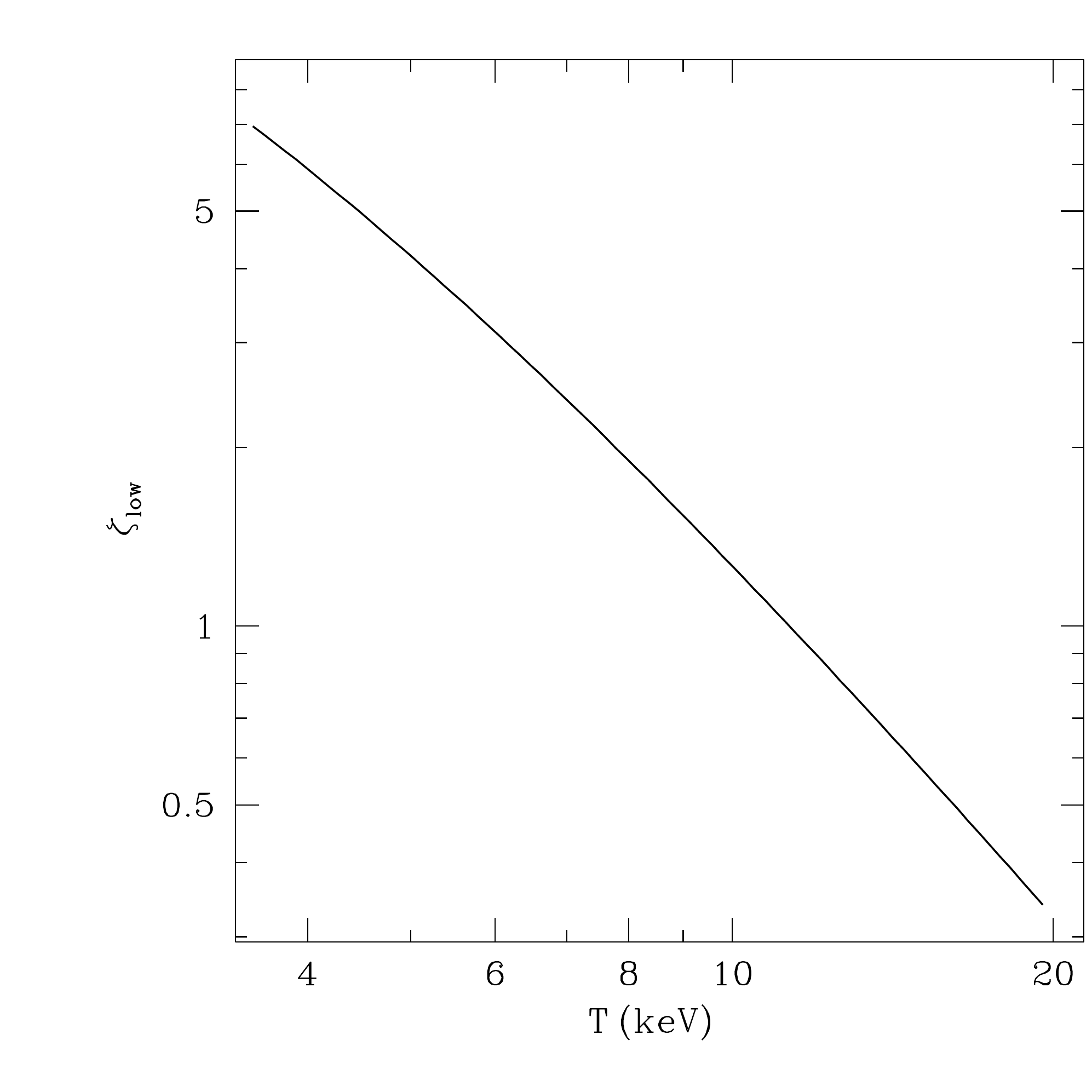}}
\caption{The dimensionless thermodynamic uncertainty factor $\zeta_\mr{low}$
as a function of maximum plasma temperature $T_\mr{max}$, calculated for
the restframe 1-5 keV bandpass. The curve is given by equation \ref{eq:zetalow}.\label{fig:zetalow} }
\vspace{0.1in}
\end{figure}}

\begin{document}

\title{Model-Independent Limits on the Line-of-Sight Depth of Clusters of
  Galaxies Using X-ray and Sunyaev-Zel'dovich Data }

\author{Andisheh Mahdavi$^{1,2}$ and Weihan Chang$^{1}$}
\affil{$^1$Department of Physics
and Astronomy, San Francisco State University, San Francisco, CA 94131
\\
$^2$Kavli Institute for Theoretical Physics, Kohn Hall,
University of California, Santa Barbara, CA 93106}

\begin{abstract}
We derive a model-independent expression for the minimum line-of-sight
extent of the hot plasma in a cluster of galaxies. The only inputs are
the 1-5 keV X-ray surface brightness and the Comptonization from
Sunyaev-Zel'dovich (SZ) data. No a priori assumptions regarding
equilibrium or geometry are required. The method applies when the
X-ray emitting material has temperatures anywhere between 0.3 keV and
20 keV and metallicities between 0 and twice solar---conditions
fulfilled by nearly all intracluster plasma. Using this method, joint
APEX-SZ and Chandra X-ray Observatory data on the Bullet Cluster yield
a lower limit of $400 \pm 56$ kpc on the half-pressure depth of the
main component, limiting it to being at least spherical, if not
cigar-shaped primarily along  the line of sight.
\end{abstract}

\keywords{Galaxies: clusters: general---cosmic microwave background---X-rays: galaxies: clusters---radiation mechanisms: thermal}

\section{Introduction}
\label{sec:introduction}

The hot intracluster medium in clusters of galaxies emits X-rays via
thermal Bremsstrahlung and line emission; the very same atmosphere
blueshifts cosmic microwave background photons via the
Sunyaev-Zel'dovich (SZ) effect \citep[for a review see][and references
  therein]{Birkinshaw99,Carlstrom02}. The X-ray emissivity scales as
electron density squared, but the SZ spectrum modification depends on
the integral of pressure along the line of sight (and hence only
linearly on the electron density). As a result, if the cosmology is
known, it is possible to recover structural information about the
cluster atmosphere \citep{Myers97,Hughes98,Grego00}. Such
``deprojection,'' however, has always required drastic assumptions
regarding the 3D structure of the ICM; in general, imposing
hydrostatic equilibrium is often required \citep{Zaroubi01}. Most
importantly, spherical or triaxial geometry have always been necessary
assumptions as well
\citep{Point02,Lee04,DeFilippis05,Ameglio07,Ameglio09,Nord09,Basu10,Allison11}. So
far, forward convolution of simple, smooth, parmeterized ICM
thermodynamic variables has been the only way to tie the X-ray and SZ
observables together to determine cluster geometry.

However, a large fraction of clusters are not relaxed and are poorly
described by hydrostatic or spherically symmetric models
\citep[e.g.][]{Eckert11}. The most spectacular and interesting of
these systems are violent mergers such as the Bullet Cluster
\citep{Clowe06}, Abell 520 \citep{Mahdavi07}, or MACSJ0025
\citep{Bradac08}. These systems clearly resist any attempt at
description using symmetric equilibrium models. At the same time, it
would be highly valuable to recover some information regarding the
structure of such objects along the line of sight. Such information
could be used to constrain the parameter space of supercomputer
N-body models that seek to replicate and thus better understand the
collisions \citep[e.g.][]{Mastropietro08,Randall08,Forero10}. 

In this Letter, we show that the Cauchy-Schwarz integral inequality
can yield useful information on the depth of a cluster of galaxies
without assumptions regarding the geometry or thermodynamic state of
the ICM. In \S2 we derive the inequality, and in \S3 we interpret
it. In \S4 we apply it to the Bullet Cluster as a test case, and in
\S5 we conclude our discussion.

\newcommand{\cool}{\Lambda_\mr{BP}(T,Z)}
\section{Theoretical Framework}

In a cluster of galaxies, the outgoing surface brightness of a column
of hot, single-phase X-ray emitting plasma is
\begin{equation}
\Sigma = \int_{E_1}^{E_2} dE \int_{-\infty}^{\infty} n_e n_H \Lambda_E(T,Z) dz 
\end{equation}
where the emission is in a rest-frame bandpass between photon energies
$E_1$ and $E_2$. The electron and hydrogen densities are $n_e$ and
$n_H$, $T$ is the plasma temperature, $Z$ is the plasma metallicity in
solar units, and $\Lambda_E$ is the cooling function incorporating the
bremsstrahlung continuum as well as line emission. For convenience we
can recast this formula as
\begin{equation}
\Sigma = \int_{-\infty}^\infty n_e^2 \cool dz
\label{eq:sigma}
\end{equation}
where $\cool$ is the cooling function integrated over the X-ray
bandpass and multiplied by the electron-to-proton ratio, $\langle
n_e/n_H\rangle$ of the plasma, which varies less than $1\%$ for
typical ICM metallicities.

In the non-relativistic limit, the fractional Sunyaev-Zel'dovich
decrement is proportional to the Compton $y$ parameter:
\begin{equation}
 y = \frac{1}{m_e c^2} \int_{-\infty}^\infty n_e \sigma_T k T dz 
\label{eq:y}
\end{equation}
where $\sigma_T$ is the Thomson cross section, $m_e$ is the electron
mass, $c$ is the speed of light, and $k$ is Boltzmann's constant. We
now also define the \emph{half-pressure depth} $z_H$ such that
\begin{equation}
 y = \frac{2}{m_e c^2} \int_{0}^{z_H} n_e \sigma_T k T dz 
\label{eq:yhalf}
\end{equation}
where the origin of the $z$ axis is arbitrary so long as the above
equation is fulfilled. In other words, half the integrated pressure of
the cluster along the line of sight is confined to a region of size
$z_H$.

The X-ray surface brightness and Compton $y$ parameter involve
integrals of the thermodynamic variables along the line of sight. They
therefore cannot be inverted without assumptions regarding the
geometry of the cluster. However, consider the Cauchy-Schwarz
inequality:

\begin{equation}
 \left[ \int f(x) g(x) dx  \right]^2 \le \int f(x)^2 dx \int g(x)^2 dx 
\end{equation}

Here we show that this integral is useful in recasting the depth
measurement problem in model-independent terms. Specifically, consider
the choice $f=n_e T$, $g = 1$; then, integrating only along the half-pressure
depth, we have
\begin{equation}
 \int_{0}^{z_H} dz \ge \frac{\left[ \int_{0}^{z_H} n_e T dz \right]^2}{\int_{0}^{z_H} n_e^2 T^2 dz}
\end{equation}

Furthermore, since the square of any function is
more compact that the function itself, equations (\ref{eq:y}) and
(\ref{eq:yhalf}) together imply that
\begin{equation}
 \int_{-\infty}^{\infty} n_e^2 T^2 dz \ge 2 \int_{0}^{z_H} n_e^2 T^2 dz
\end{equation}
The above inequality follows from the fact that the variance of any
square-integrable function $f(x)$ is equal to or larger than the
variance of $f(x)^2$. Therefore it is always possible to choose a
z-axis origin such that $[0,z_H]$ contains half the integrated
pressure but more than half the integral of the pressure squared. We
therefore obtain

\begin{equation}
 \int_{0}^{z_H} dz \ge 2 \frac{\left[ \int_{0}^{z_H} n_e T dz \right]^2}{\int_{-\infty}^{\infty} n_e^2 T^2 dz}
\label{eq:dz}
\end{equation}


Combining equation (\ref{eq:dz}) with the Comptonization parameter (equation
\ref{eq:y}), and multiplying by $\Sigma/\Sigma = 1$, we can write
\begin{equation}
z_H \ge  \frac{m_e^2 c^4 y^2}{2 \sigma_T^2 \int_{-\infty}^\infty n_e^2 T^2 dz} \times \frac{\Sigma}{\Sigma}
\end{equation}
This constraint on cluster depth can be rewritten as
\newcommand{\zt}{\zeta}
\begin{equation}
 z_H \ge \ell \times \zt
\label{eq:zh}
\end{equation}
where we define $\ell$ as the fiducial depth of the cluster
\begin{eqnarray}
\ell & \equiv & \frac{m_e c^3 y^2}{2 \sigma_T \Sigma}\\
 & \approx & 0.56 \left(\frac{y}{10^{-4}} \right)^2 \left(\frac{\Sigma}{10^{45} \mr{\ erg\ } \mr{s}^{-1} \mr{\ Mpc}^{-2}} \right)^{-1} \mr{Mpc}
\label{eq:ell}
\end{eqnarray}
and where we define the dimensionless, bandpass-dependent thermodynamic uncertainty factor 
\begin{equation}
 \zt \equiv \frac{m_e c\int n_e^2 \cool dz}{\sigma_T \int n_e^2 k^2 T^2 dz } 
\label{eq:zeta}
\end{equation}
If we were dealing with a medium of constant temperature and
metallicity, we could now pull out the functions of $T$ and $Z$,
cancel the emission measure integrals, and be done; but we do not make
this assumption and continue with the general case.

So far we have shown that given a measurement of the Compton $y$
parameter and the X-ray surface brightness, the half-pressure depth of
the cluster along the line of sight is at least $\ell$ (equation
\ref{eq:ell}) times a correction factor $\zt$ (equation
\ref{eq:zeta}).

We now attempt to derive useful bounds on $\zt$. Suppose that, for any
given X-ray bandpass, $\zt > \zeta_\mr{low}$ for all plausible values
of $T$ and $Z$. In that case without any loss of generality we can
write
\begin{equation}
z_H \ge \ell \times \zeta_\mr{low}
\end{equation}
To find $\zeta_\mr{low}$, imagine three arbitrary one dimensional
positive functions $n_e(z)$, $p(z)$ and $q(z)$. If $p/q >
\zeta_\mr{low} $ everywhere, then necessarily $p > \zeta_\mr{low} q$,
and so $n_e^2 p > \zeta_\mr{low} n_e^2 q$ everywhere. It follows that
$\int_a^b n_e(z)^2 p(z)~>~\zeta_\mr{low} \int_a^b n_e(z)^2 q(z)$. In
other words, the ratio of two one-dimensional integrals is always
greater than the smallest value taken on by the ratio of the
integrands, as long as the integrands are positive and the
limits of integration are the same.

Choosing
$p=m_e c \cool$ and $q = \sigma_T k^2 T^2$, it follows that $\zt >
\zeta_{low}$ if
\begin{equation}
\zeta_\mr{low} = \mr{minimum\ of\ } \frac{m_e c \cool}{\sigma_T k^2 T^2}
\end{equation}
for all $Z$ and $T$ of interest. Our task then becomes one of 
minimizing the quantity
\begin{equation}
\gamma = \frac{m_e c \cool}{\sigma_T k^2 T^2}.
\end{equation}

\section{Interpretation}

The lower limit on the half-pressure depth of a cluster of galaxies
becomes $\ell \times \zeta_\mr{low}$, where $\zeta_\mr{low}$ is the
minimum of the bandpass-dependent quantity $\gamma$ over all
temperatures and metallicities contributing to the SZ and X-ray
fluxes.  Since the cooling function for thermal Bremsstrahlung
$\Lambda(T,Z) \sim \sqrt{T}$ for $T \gg 10$ keV, it follows that
$\gamma~\sim~T^{-3/2}$ for large $T$, and the $z_H$ constraint
requires the assumption of a high-temperature cutoff in order to be
relevant (i.e., to avoid the trivial result $z_H > 0$). We call this
high-temperature cutoff $T_\mr{max}$. We leave $T_\mr{max}$ flexible
on a cluster-by-cluster basis. Its value can be constrained via
modeling of sufficiently high signal-to-noise X-ray spectra, or else
set to a sufficiently high value (e.g. no cluster is believed to
possess thermal gas beyond 20 keV).

A low temperature cutoff, $T_\mr{min}$, is also needed, because
$\zeta_\mr{low} \rightarrow 0$ whenever $\cool \rightarrow 0$, i.e.,
when there is a column of material emitting outside the X-ray
bandpass. In a bandpass centered on photon energy E, all emission with
$T \ga E/3$ contributes to the X-ray flux, but emission with $T \la
E/3$ contributes negligibly.  For most common X-ray bandpasses, this
translates to $T < 0.3$ keV. Searches for very cool, $T< 0.3$ keV
material---typically referred to as the warm-hot intergalactic medium
or WHIM---have thus far been inconclusive. While the WHIM must exist,
it likely has densities $\la 5 \times 10^{-5}$ cm$^{-3}$
\citep{Nicastro05}, and thus would contribute negligibly to the
overall SZ and X-ray flux in the direction of a massive cluster, where
the typical gas densities are $10^{-4}$ cm$^{-3}$ or higher.  We
therefore assume that in any source with both a significant SZ
decrement and a measured Chandra/XMM-Newton X-ray flux, the
contribution of material at $T < 0.3$ keV is negligible. For example,
consider a medium at $T = 0.3$ keV, $n_e = 5 \times 10^{-5}$ cm$^{-3}$,
emitting over a length of 10 Mpc along the line of sight. Such a
medium would make an additive contribution of $6 \times 10^{-7}$ to
the central Compton $y$ parameter, a less than one percent effect for
massive clusters even in the maximal WHIM case.  We therefore neglect
the WHIM for our purposes and set $T_\mr{min} = 0.3$ keV.

As the bandpass to focus on in the present Letter, we select the
restframe 1-5 keV photon energy range. This energy range has the
advantage of overlapping the most sensitive regions of current X-ray
telescopes, and of being detectable between redshifts $z=0$ and $z=1$
with relatively low astrophysical background contamination. 

With the bandpass chosen, we numerically minimize $\gamma$ as a
function of temperature and metallicity, assuming a MEKAL
\citep{Mewe85} plasma with allowed metallicities from 0 to twice
solar, and allowed temperatures ranging from $T_\mr{min}$ to
$T_\mr{max}$.  We find that the minimum value of $\gamma$ is
sensitive only to $T_\mr{max}$. For all $T_\mr{min} > 0.3$ keV,
$T_\mr{max} > 3.5$ keV we find that the following empirical formula
fits the exact value of $\zeta_\mr{low}$ with better than 1\%
accuracy:
\begin{equation}
\zeta_\mr{low} = 1.883 + \frac{54.12}{T_\mr{max} / keV}  - \frac{19.05}{\left(T_\mr{max} / keV \right)^\frac{1}{2}}.
\label{eq:zetalow}
\end{equation}
I.e. $\zeta_\mr{low}$ has no dependence on the coolest material so
long as its temperature is above 0.3 keV. We note that beyond $T=20$
keV the relativistic correction to equation (\ref{eq:y}), which we
neglect here, becomes important. The above formula is therefore only
valid between $T_\mr{max} = 3.5$ and $20$ keV.


\section{Application to the Bullet Cluster}

To demonstrate this technique, we jointly analyze X-ray and
Sunyaev-Zel'dovich observations of the Bullet Cluster. This cluster is
a particularly apt candidate for this type of analysis, due to the
clear incorrectness of typical spherically symmetric models in
describing its morphology. The Bullet Cluster consists of a ``main''
diffuse component plus a ``bullet'' to the west. Using 150GHz APEX-SZ,
\cite{Halverson09} infer a Comptonization for the main component of
the cluster of $y_0 = 3.4 \pm 0.3 \times 10^{-4}$. This Compton $y$
measurement is not model-independent---it relies to some extent on the
precise 2D shape of the cluster in the plane of the sky, reflecting
the limited angular resolution of APEX-SZ experiment.  Further
uncertainties arise from the fact that some leakage from the Bullet
1.5\arcmin\ to the west into the central APEX-SZ beam is likely. It is
difficult to estimate the bias in $y$ introduced here, first because
the Bullet is both denser and cooler than the main component of the
cluster, and second because the Compton $y$ model adopts the X-ray
profile as a strong prior. A further issue is that the azimuthally
averaged profile has better statistics at radii which include the
Bullet (and therefore the fit may be particularly driven by those
regions).  These shortcomings highlight the need for higher angular
resolution SZ experiments if bias-free 3D measurements of cluster
structure are to be achieved. Such observations would allow for more
complicated 2D Compton $y$ distributions and thus overcome much of the
systematic problems of the smooth symmetric models. 

To measure the rest-frame 1-5 keV surface brightness, we reanalyze an
archived 100ks Chandra X-ray Observatory data set (ObsID 5356). We
extract a region of radius 182 kpc (0.691$^\prime$ at $H_0 =
71\ \kms\ \Mpc$\m, $\Omega_M=0.23$) centered on the main portion
(i.e., the true X-ray center of the diffuse component of the
cluster). The aperture was selected to match the beam size of the
APEX-SZ experiment. We refer to \cite{Mahdavi07} for all X-ray data
reduction procedures, including particle background subtraction, point
source masking, and other details.

Using a redshift $z=0.296$, we obtain a best-fit temperature of $11.5
\pm 0.6$ keV, best-fit metallicity of $0.23 \pm 0.05$ solar, and
best-fit absorbing column of $7.1 \pm 0.5 \times 10^{20}$
cm$^{-2}$. The X-ray flux in the 0.77-3.9 keV band (corresponding to
the 1-5 keV rest frame band) is $1.89 \pm 0.01 \times 10^{-12}$ ergs
cm$^{-2}$ s\m. Multiplying by $4 \pi$ times the square of the
luminosity distance at the above cosmology (1517 Mpc), and dividing by
the projected area of 0.104 Mpc$^2$, we obtain a restframe surface
brightness of $5.03 \pm 0.03 \times 10^{45}$ ergs s\m Mpc$^{-2}$.

Via equation (\ref{eq:ell}) the observations therefore imply an upper
limit
\begin{equation}
z_H \ge \left[ 0.56 \times (3.2)^2 \times (5.0)^{-1} \right] \zeta_\mr{low}
\label{eq:ellbullet}
\end{equation}
To constrain $\zeta_\mr{low}$, we need an estimate of the highest
temperature gas along the line of sight. To be conservative, we assume
that temperatures up to $20$ keV contribute significantly to the
emission measure (this is consistent with the measured X-ray
spectrum). Plugging $20$ keV into equations \ref{eq:zetalow} and
\ref{eq:ellbullet} yields a lower limit
\begin{equation}
z_H \ge 400 \pm 56\ \mr{kpc}
\end{equation}
where the dominant source of error is the statistical uncertainty on
the Compton $y$ parameter. The (unknown) systematic bias in $y$ is not
included.

Now we attempt to calculate the axial ratio of the main (eastern)
component of the cluster, comparing the line-of-sight half-pressure
depth $z_H$ to the plane-of-sky half-pressure width, which we call
$w_H$.  \cite{Halverson09} find that
\begin{equation}
\Delta T /T \propto (1+\theta^2/\theta_c^2)^{1-3 \beta/2}
\end{equation}
 where $\beta= 1.2 \pm 0.13$ and $\theta_c = 142\arcsec \pm
 18\arcsec$.  Numerically integrating this profile, we find that half
 the flux is contained within a circular aperture of radius $0.67
 \theta_c$, which translates into a half-width width of $w_H \approx 2
 \times 0.67 \times 325$ kpc $ \approx 435$ kpc. Note that this is
 likely more an upper limit than an accurate measurement, given the
 mixing of the bullet and main component signals due to poor
 resolution of the APEX-SZ instrument.  However, the
 direction of this inequality works in our favor, because a
 lower limit on $z_H$ divided by an upper limit on $w_H$ still
 yields a valid lower limit on $z_H/w_H$:

\begin{equation}
\frac{z_H}{w_H} \ge 0.92
\end{equation}

Thus, the ratio of the half-pressure diameter of the Bullet Cluster to
its half-pressure width in the sky is constrained to be greater than
approximately one.  

\fone



\section{Conclusion}

We derive the first completely model- and geometry-independent
expression limiting the half-pressure depth of a galaxy cluster along
the line of sight. Regardless of geometry, clumping, or thermodynamic
state of the intracluster medium, combining X-ray and
Sunyaev-Zel'dovich observations of a single cluster can yield a lower
limit on this depth. This lower limit is
\begin{equation}
z_H \ge \frac{m_e c^3 y^2}{2 \sigma_T \Sigma} \zeta_\mr{low}
\end{equation}
where $\zeta_\mr{low}$ is a dimensionless number which depends on the
specific bandpass used and the maximum ICM temperature that
contributes significantly to the SZ effect. Equation
(\ref{eq:zetalow}) gives an empirical fitting formula for
$\zeta_\mr{low}$ in the 1-5 keV restframe bandpass (also shown in Figure 1).

This lower limit is a function of largely model-independent observables:
$y$ (the Compton $y$ parameter), and $\Sigma$ (the X-ray surface
brightness). No deprojection, and no modeling beyond standard
reduction of the SZ decrement and X-ray spectrum is required. The
method neglects relativistic SZ corrections, and assumes that the bulk
of the pressure and cooling is done by gas with temperatures between
0.3 keV and 20 keV, and metallicities between 0 and 2 times
solar. Otherwise, the gas can have an arbitrary distribution along the
line of sight. The constraint can be made more stringent by relaxing
the upper limit of gas temperatures allowed; one way to do this in the
future is via measurements of the differential emission measure $d n_e^2
/ dT$. This requires some more sophisticated spectral modeling, but
still no assumptions regarding geometry or equilibrium.

These results are useful in a number of scenarios. For merging
clusters of galaxies such as the Bullet Cluster, they can help
constrain the geometry of the merger. Using this method, our analysis
of joint APEX-SZ and Chandra data for the Bullet Cluster imply a
minimum line-of-sight half-pressure depth of $>400$ kpc, implying a
minimum line-of-sight to plane of the sky axial ratio of $\approx
1$.

Thus the Bullet Cluster is constrained to be at least spherical, if
not cigar-shaped primarily along  the line of sight. For larger
samples of X-ray emitting clusters, this technique can help quantify
possible selection biases towards line-of-sight elongations, as well
as possibly aid in determining the geometry of shock and cold fronts.

Wee thank James Allison and the anonymous referee for useful
comments. This research was made possible by NASA through Chandra
award No. AR0-11016A, issued by the Chandra X-ray Observatory Center,
which is operated by the Smithsonian Astrophysical Observatory for and
on behalf of NASA under contract NAS8-03060. This research was
supported in part by the National Science Foundation under Grant
No. NSF PHY05-51164.

\vspace{0.2in}

\ifthenelse{\isundefined{\usemyrefs}}{
\section*{References}
\input{ms.bbl}
}{ \bibliographystyle{myrefs/apj}
\bibliography{myrefs/myrefs} }

\end{document}